\documentclass[prl,twocolumn,showpacs,amsmath,amssymb,citeautoscript]{revtex4}

 \def\be{\begin{eqnarray}}
\def\ee{\end{eqnarray}}

\usepackage{graphicx}

\begin{document}

\title{Effects of the anomaly on the two-flavor QCD chiral phase transition}
\author{Shailesh Chandrasekharan and Abhijit C. Mehta}
\affiliation{ Department of Physics, Box 90305, Duke University,
Durham, North Carolina 27708.}

\begin{abstract}
We use strongly coupled lattice QED with two flavors of massless staggered fermions to model the physics of pions in two-flavor massless QCD. Our model has the right chiral symmetries and can be studied efficiently with cluster algorithms. In particular we can tune the strength of the QCD anomaly and thus study its effects on the chiral phase transition. Our study confirms the widely accepted view point that the chiral phase transition is first order in the absence of the anomaly. Turning on the anomaly weakens the transition and turns it second order at a critical anomaly strength. The anomaly strength at the tricritical point is characterized using $r = (M_{\eta'}-M_{\pi})/\rho_{\eta'}$ where $M_{\eta'}, M_{\pi}$ are the screening masses of the anomalous and regular pions and $\rho_{\eta'}$ is the mass-scale that governs the low energy fluctuations of the anomalous symmetry. We estimate that $r \sim 7 $ in our model. This suggests that a strong anomaly at the two-flavor QCD chiral phase transition is necessary to wash out the first order transition.
\end{abstract}

\maketitle

Understanding the nature of the finite temperature chiral phase transition in QCD has been an area of active research for almost two decades. The subject is important for the field of relativistic heavy ion experiments \cite{Brown:2003wa}. It also plays a role in cosmology where a first order transition can lead to rich cosmological scenarios \cite{PhysRevD.30.272}. The conventional wisdom regarding the nature of the chiral phase transition in $N_f$-flavor QCD comes from the  renormalization group study of linear sigma models which describe the fluctuations of the order parameter in QCD. These models are invariant under $SU(N_f)\times SU(N_f) \times U(1)$ chiral symmetry. The anomaly is introduced through a term that breaks the $U(1)$ symmetry explicitly. These studies, usually based on the $\epsilon$-expansion reveal that the number of light quark flavors and the strength of the anomaly can play a significant role in determining the order of the transition \cite{Pisarski:1983ms}. While a second order phase transition is possible for $N_f=2$, the presence of more light quarks can introduce fluctuations that will force the transition to become first order. Similarly, if the anomaly is sufficiently weak at the transition, the two-flavor transition could itself turn into a first order one. It is known that the anomaly weakens with increase in temperature and number of colors. What really occurs in nature remains a topic of active research even today.\cite{D'Elia:2005sy,Engels:2005rr,deForcrand:2006pv}. Recent reviews can be found in \cite{Laermann:2003cv,Heller:2006ub}.

Although the $\epsilon$-expansion is known to be unreliable in three dimensions, Monte Carlo methods have confirmed some of its predictions. Both non-perturbative studies in the sigma models and lattice QCD calculations have found that for three or more flavors of light quarks the chiral transition is first order as predicted. Further, there is little doubt that in the presence of a large anomaly, the two flavor transition can be second order in the three dimensional $SU(2)\times SU(2) \sim O(4)$ universality class \cite{ParisenToldin:2003hq}. On the other hand, the effects of a weak anomaly remains relatively unexplored. Studies within mean field theory, not surprisingly, agree with the predictions of the $\epsilon$-expansion, which forbids the two-flavor transition without the anomaly to be second order \cite{Lenaghan:2000kr,Marchi:2003wq}. However, recent renormalization group studies, based on three dimensional perturbation theory and resummation techniques, suggest that a second order phase transition is indeed possible in the absence of the anomaly \cite{Basile:2005hw}. It would be interesting to identify and study this new, $SU(2)\times SU(2) \times U(1) \sim O(4)\times O(2)$ second order critical behavior using Monte-Carlo methods if it exists. To do this, one will have to study  $O(4)\times O(2)$ symmetric sigma models with the relevant symmetry breaking pattern and look for the predicted continuous transition. As far as we know this is difficult in the conventional formulation of the sigma models due to the lack of efficient algorithms and hence remains unexplored.

Here we offer a new approach to the subject. Using strongly coupled lattice QED with two flavors of staggered fermions, which can be studied very efficiently with recently discovered cluster algorithms \cite{Adams:2003cc,Chandrasekharan:2006tz,Cecile:2006}, we model two-flavor massless QCD in the absence of the anomaly. Our model has the right symmetries and symmetry breaking pattern and so should be viewed as an alternative formulation of the relevant $O(4)\times O(2)$ sigma model. An additional four-fermion term can be used to introduce the effects of the anomaly and this allows us to efficiently study the phase diagram as a function of temperature and anomaly strength using Monte Carlo methods for the first time. Although our study is not an exhaustive search for the second order critical point predicted in \cite{Basile:2005hw}, it can be viewed as a first attempt within the context of a specific model. 

The action of our model is given by
\begin{eqnarray}
S &=& - \sum_{x}\sum_{\mu=1}^{4}
\eta_{\mu,x}\bigg[\mathrm{e}^{i\phi_{\mu,x}}{\overline\psi}_x
{\psi}_{x+\hat\mu}
-\mathrm{e}^{-i\phi_{\mu,x}}{\overline\psi}_{x+\hat\mu}{\psi}_x\bigg]
\nonumber \\
&& -\sum_x \frac{C}{2}\bigg({\overline\psi}_x{\psi}_x \bigg)^2,
\label{eq1}
\end{eqnarray}
where $x=(\vec{x},t)$ denotes a lattice site on a $L^3 \times 4$ hypercubic lattice. $\overline\psi_x$ and $\psi_x$ are two component Grassmann fields that represent $(u,d)$ quarks, and $\phi_{\mu,x}$ is the compact $U(1)$ gauge field through which the quarks interact. Here $\mu = 4$ denotes the temperature direction,while $\mu=1,2,3$ are the spatial directions. The staggered fermion phase factors $\eta_{\mu,x}$ obey the relations: $\eta_{4,x}^2 = T$ and $\eta_{i,x}^2 = 1$ for $i=1,2,3$. The parameter $T$ controls the fictitious temperature. The coupling $C$ sets the strength of the anomaly. 

The above action exhibits a global $SU_L(2)\times SU_R(2)$ symmetry. Indeed it is easy to check that the action is invariant under
$\psi_{x_e} \rightarrow L\psi_{x_e}, \psi_{x_o} \rightarrow R\psi_{x_o}, 
\overline{\psi}_{x_o} \rightarrow \overline{\psi}_{x_o} L^\dagger, 
\overline{\psi}_{x_e} \rightarrow \overline{\psi}_{x_e} R^\dagger$, where $L,R\in SU(2)$ and $x_e$ and $x_o$ refer to even and odd sites. When $C=0$ the action is also invariant under a $U(1)$ transformation given by $\psi_{x_e} \rightarrow \mathrm{e}^{i\theta}\psi_{x_e}, \psi_{x_o} \rightarrow \mathrm{e}^{-i\theta}\psi_{x_o}, \overline{\psi}_{x_o} \rightarrow \overline{\psi}_{x_o}\mathrm{e}^{-i\theta}, \overline{\psi}_{x_e} \rightarrow \overline{\psi}_{x_e}\mathrm{e}^{i\theta}$. This will be interpreted as the anomalous $U_A(1)$ transformation of QCD. When $C\neq 0$, the $U_A(1)$ is explicitly broken to $Z_2$ and the action is invariant only under $SU_L(2) \times SU_R(2) \times Z_2$.  Thus, our model has the same symmetries as $N_f=2$ QCD with a parameter $C$ which helps change the anomaly strength.

\begin{figure}[htb]
\begin{center}
\includegraphics[width=7cm]{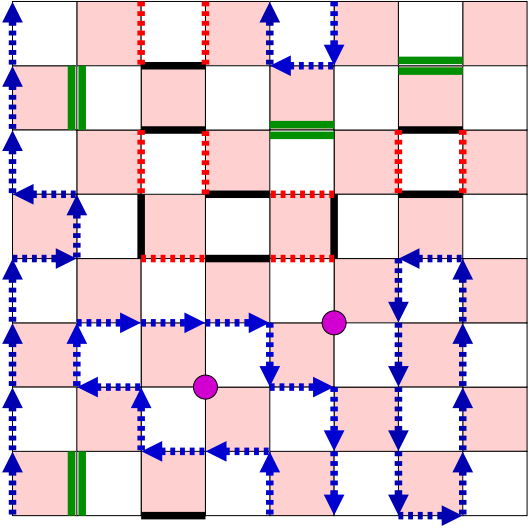}
\end{center}
\caption{\label{fig1}An example of a DPI configuration in two dimensions.}
\end{figure}

The partition function of our model can be expressed as a classical statistical mechanics problem involving gauge invariant objects: dimers (D), pion loops (P) and instantons (I)\cite{Cecile:2006,Chandrasekharan:2006zq,Cecile:2006zr}. We refer to the set of these configurations as DPI configurations and denote them $\{{\cal K}\}$. Each DPI configuration is characterized by a site variable $I(x)\in \{0,1\}$ representing instantons and three bond variables, $\pi^u_\mu(x)\in \{0,1\}$, $\pi^d_\mu(x) \in \{0,1\}$, $\pi^1_\mu(x)\in\{-1,0,1\}$, which give rise to dimers and pion loops. In our notation $\pi^u_{-\mu}(x) = \pi^u_\mu(x-\hat{\mu})$ and similarly for other bond variables. Due to the Grassmann nature of the fermion fields the following constraints must also be satisfied at each site $x$:
$\sum_{\mu} \pi_{\mu}^1(x) = 0$, $2I(x) + \sum_{\mu} [\pi_{\mu}^u(x) + \pi_{\mu}^d(x) + |\pi_{\mu}^1(x)|] = 2$, $\sum_{\mu}[\pi_{\mu}^u(x) - \pi_{\mu}^d(x)] = 0$,
where the sum over $\mu$ goes over $\pm 1,...\pm 4$. Figure \ref{fig1} gives an illustration of a DPI configuration in $1+1$ dimensions. The partition function is given by $Z = \sum_{[{\cal K}]} \mathrm{e}^{-S}$ where $S = - \sum_x [\pi_4^u(x) +\pi_4^d(x) + \pi_4^1(x)]\log(T) + I(x)\log(C)$. Although the $SU(2)\times SU(2)\times U_A(1)$ symmetry of the model is not apparent in the DPI formulation, it is easy to check that $J^V_\mu(x) = \pi^1_\mu(x)$, $J^C_\mu(x) =  \varepsilon(x) [\pi^u_\mu(x) - \pi^d_\mu(x)]$ and $J^A_\mu(x) = \varepsilon(x) [\pi_i^u(x) + \pi_i^d(x) + |\pi_i^1(x)|-1/2]$ form the vector, chiral and axial conserved currents when $C=0$. Note $\varepsilon(x)$ is defined to be $+1$ on even sites and $-1$ on odd sites.

We will focus on four observables:
\begin{itemize}
\item[(1)] Vector current susceptibility
\begin{equation}
Y_C = \frac{1}{3L^3}\Bigg\langle \sum_{i=1}^3 \Bigg(\sum_x J^C_i(x) \Bigg)^2\Bigg\rangle.
\end{equation}
\item[(2)] Axial current susceptibility
\begin{equation}
Y_A = \frac{1}{3L^3}\Bigg\langle \sum_{i=1}^3 \Bigg(\sum_x J^A_i(x)\Bigg)^2\Bigg\rangle.
\end{equation}
\item[(3)] Chiral condensate susceptibility
\begin{equation}
\chi = \frac{1}{4L^3}\Bigg\langle \Bigg(\sum_{x} \overline{\psi}(x)\psi(x)\Bigg)^2\Bigg\rangle.
\end{equation}
\item[(4)] Mass of the anomalous pion $M_{\eta'}$ obtained from the zero momentum correlation function
\begin{equation}
G(x)=\frac{1}{4 L^2}\sum_{x_\perp,y_\perp} 
\langle \varepsilon\overline{\psi}\psi(x,x_\perp) 
\ \varepsilon\overline{\psi}\psi(0,y_\perp) \rangle,
\end{equation}
by fitting it to the form $A \cosh(M_\eta' x)$. Here $x_\perp$ and $y_\perp$ represent coordinates perpendicular to the direction $x$. 
\end{itemize}

\begin{figure}
\begin{center}
\includegraphics[width=0.45\textwidth]{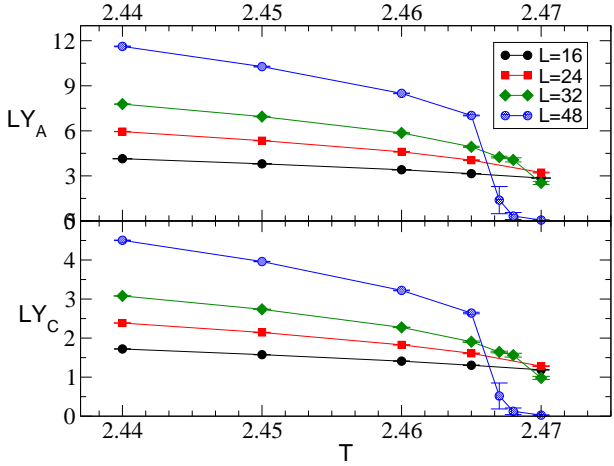}
\end{center}
\caption{\label{fig2} Plot of $LY_A$ and $LY_C$ versus $T$ for different values of $L$ for $C=0$. The lack of a point where all the curves cross shows the absence of a second order transition. We can estimate $T_c\sim 2.466(1)$ }
\end{figure}

At $C=0$ and small $T$ we expect that the $SU_L(2)\times SU_L(2)\times U_A(1)$ symmetry of the model is broken to a diagonal $SU(2)$ flavor symmetry. As $T$ is increased one expects a phase transition at $T=T_c$ where symmetries are restored. This phase transition can be studied conveniently using the two spin-stiffnesses defined as $\rho_\pi = \lim_{L\rightarrow \infty} Y_C$ and $\rho_{\eta'} = \lim_{L\rightarrow \infty} Y_A$. We expect these two quantities to be non-zero in the broken phase and to vanish in the symmetric phase. With knowledge of these, one can write down the leading term of the low energy effective action that governs the Goldstone-boson fluctuations in the broken phase. This action turns out to be
\begin{equation}
S =  \int d^3 x \Bigg\{\frac{\rho_\pi}{4}
\mathrm{Tr}\Big[\partial_\mu U^\dagger \partial_\mu U\Big] +
\frac{\rho_{\eta'}}{2}|\partial_\mu u|^2 \Bigg\}
\end{equation}
with $U(x)\in SU(2)$ and $u(x) = \exp(i\eta')$. Clearly $\rho_\pi$ and $\rho_{\eta'}$ define mass-scales for the problem. If the transition to the symmetric phase is second order, we expect $\rho \sim A (T_c-T)^\nu$. We also expect the finite size scaling relation:
\begin{equation}
\label{fss}
L Y(L) \sim \sum_{k=0}^3 f_k \Big[(T_c-T) L^{1/\nu}\Big]^k 
\end{equation}
in the region $(T_c-T)L^{1/\nu} \ll 1$. In other words a second order transition will be characterized by the fact that $L Y(L)$ is independent of the lattice size at $T=T_c$. We will use this characteristic to decide if a transition is second order or not. In Figure \ref{fig2} we plot $LY_A(L)$ and $LY_C(L)$ as a function of $T$ at $C=0$. Although there is clear evidence for a transition at $T_c \approx 2.466(1)$, it is not second order. This result confirms the $\epsilon$-expansion scenario.

\begin{figure}
\begin{center}
\includegraphics[width=0.45\textwidth]{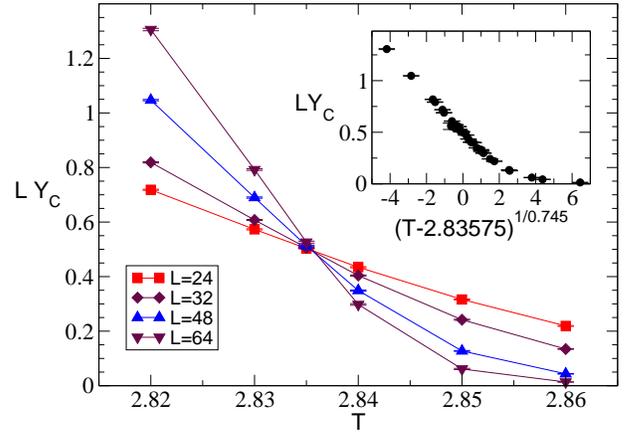}
\end{center}
\caption{\label{fig3} Plot of $LY_C$ versus $T$ for different values of $L$ for $C=0.3$. The presence of a point where all the curves cross shows the presence of a second order transition. We can estimate $T_c = 2.83555(10)$. The inset shows that all our data fits $O(4)$ scaling extremely well.}
\end{figure}

As discussed in the introduction, when the anomaly is large one expects $O(4)$ critical behavior. In order to test this we set $C=0.3$. In this case we expect critical behavior only in $Y_C(L)$ since the anomalous pion $\eta'$, will remain massive. In Figure \ref{fig3} we plot $L Y_C(L)$ as a function of $T$. Unlike the $C=0$ case, now all the curves cross at a point. We can fit all our data to the 3d $O(4)$ second order scaling form ($\nu=0.745$ \cite{ParisenToldin:2003hq}) given in eq.(\ref{fss}) with a $\chi^2/DOF=1.3$. The fit yields $f_0=0.494(2)$, $f_1=0.388(4)$, $f_2=0.022(4)$, $f_3=-0.008(2)$, $T_c=2.83555(10)$. The scaled data is shown in the inset of Figure \ref{fig3}.

\begin{figure}
\vskip0.3in
\begin{center}
\includegraphics[width=0.45\textwidth]{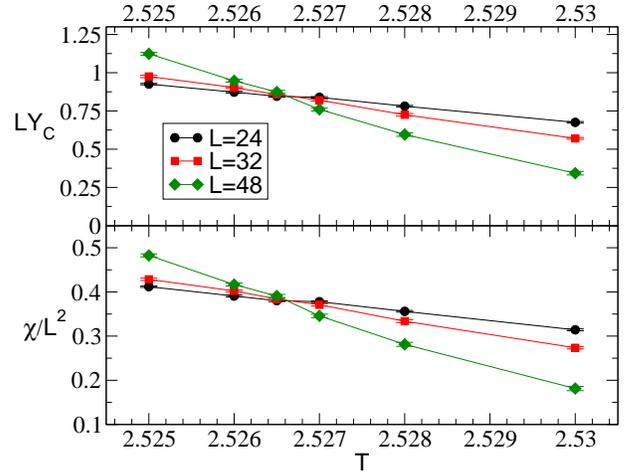}
\end{center}
\caption{\label{fig4} Plot of $LY_C$ (top) and $\chi/L^2$ (bottom) versus $T$ for different values of $L$ for $C=0.03$. The scaling of $\chi$ assumes $\eta=0$. The presence of a point where all the curves cross shows the presence of a second order transition. We can estimate $T_c\sim 2.52652(3)$.}
\end{figure}

\begin{figure}
\begin{center}
\includegraphics[width=0.45\textwidth]{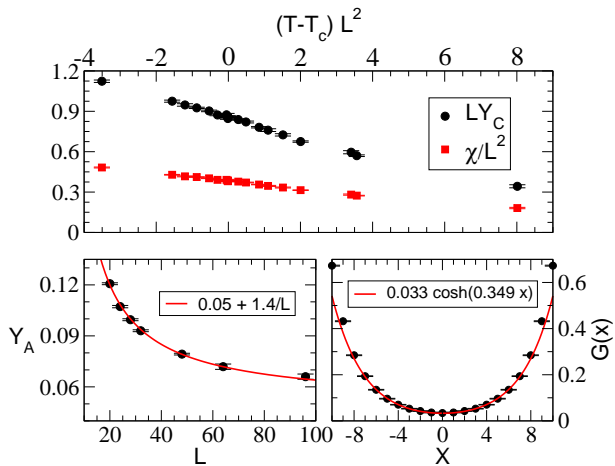}
\end{center}
\caption{\label{fig5} The plots shown are close to the tricritical point $C=0.03$ and $T=2.5265$. Assuming $\nu=0.5$ and $\eta=0$, in the top panel we show  $LY_C$ and $\chi/L^{2-\eta}$ as a function of $(T-T_c) L^{1/\nu}$ for all of our data. The fact that all the data fall on a single curve demonstrates mean field universality. The bottom left panel is the plot of $Y_A$ versus $L$ and the right panel shows the correlation function $G(x)$. Solid lines are fits discussed in the text.}
\end{figure}

Given that the transition is first order at $C=0$ and second order at $C=0.3$ it is interesting to locate the tricritical point. After some trial and error we have found that the tricritical point is located at $C\approx 0.030(5)$ and $T_c\approx 2.52652(3)$. At a tricritical point one expects mean field scaling in three dimensions \cite{Amit}, i,e.. $\nu=0.5$ and $\eta=0$. In addition to eq.~(\ref{fss}), here we have also looked at the critical scaling relation for the chiral condensate susceptibility
\begin{equation}
\label{fss1}
\chi \approx L^{2-\eta} \sum_{k=0}^3 g_k \Big[(T_c-T) L^{1/\nu}\Big]^k.
\end{equation}
In Figure \ref{fig4} we show the behavior of $LY_C$ and $\chi/L^2$ as a function of $T$ for $C=0.03$.  Clearly, again the curves meet at a point. A combined fit of all our data to eqs.~(\ref{fss},\ref{fss1}), assuming mean field scaling, gives a $\chi^2/DOF = 1.1$. The fit yields $f_0=0.853(3)$, $f_1=0.091(2)$, $g_0=0.384(1)$, $g_1=0.078(2)$, $T_c=2.52652(3)$. In the fit, $f_k,g_k$ for $k\geq 2$ were set to zero. In the top plot of Figure \ref{fig5}, we show the scaled observables.

It is useful to characterize the strength of the anomaly along the critical line from $C=0$ to $C=0.03$ using a dimensionless parameter. In order to do this we consider two mass scales naturally defined along this line. The first mass scale is $\rho_{\eta'} \equiv [(\rho_{\eta'})_{T_c^-} + (\rho_{\eta'})_{T_c^+}]/2$ which is the average value of the anomalous spin stiffness across the transition. The second mass scale is the difference between the screening masses of the anomalous pion $M_{\eta'}$ and the regular pion $M_\pi$. The dimensionless ratio $r = (M_{\eta'}-M_\pi)/\rho_{\eta'}$ clearly measures the strength of the anomaly: It is zero at C=0, i.e., in the absence of the anomaly, and increases with $C$. In Figure \ref{fig5} (bottom-left) we show the finite size scaling of $Y_A(L)$ close to the tricritical point. By fitting the data to the form $\rho_\eta'+a/L$ we find that $\rho_\eta'\approx 0.05$. Note $(\rho_{\eta'})_{T_c^-} =(\rho_{\eta'})_{T_c^-}$ at the tricritical point. We have noticed that $\rho_{\eta'}$ does not change much from $C=0$ to $C=0.03$ along the critical line. This suggests that $r$ is indeed a good parameter to measure the strength of the anomaly. In Figure \ref{fig5} (bottom-right) we show the correlation function $G(x)$ again close to the tricritical point, for $L=20$. By fitting the data to $A\cosh(M_\eta' x)$ we estimate $M_\eta' \approx 0.35$. Since $M_\pi = 0$ at the tricitical point, we conclude that $r \approx 7$. Thus, $r$ changes from zero in the absence of the anomaly to approximately $7$ at the tricritical point.

Although $r$ is an interesting ratio, it is clearly non-universal since it involves non-critical scales. So we cannot predict the value of $r$ in QCD from our results. However, the large value found here at the tricritical point may be a generic feature of many theories including QCD. Thus, a strong anomaly may be necessary before the second order $O(4)$ critical behavior sets in. On the other hand lattice QCD calculations suggest that the anomaly is weak at the chiral phase transition \cite{Chandrasekharan:1998yx}. If this is true a second order chiral transition in the $O(4)$ universality class for two-flavors of massless quarks in QCD seems unlikely.

We thank E.Vicari and Uwe-Jens Wiese for helpful comments. This work was supported in part by the Department of Energy grant DE-FG02-05ER41368. Preliminary versions of this work can be found in the undergraduate senior thesis of ACM and in a conference proceedings \cite{Chandrasekharan:2006zq}. We thank Bern university, where this work was completed, for hospitality.

\bibliography{qcd}

\begin{thebibliography}{19}
\expandafter\ifx\csname natexlab\endcsname\relax\def\natexlab#1{#1}\fi
\expandafter\ifx\csname bibnamefont\endcsname\relax
  \def\bibnamefont#1{#1}\fi
\expandafter\ifx\csname bibfnamefont\endcsname\relax
  \def\bibfnamefont#1{#1}\fi
\expandafter\ifx\csname citenamefont\endcsname\relax
  \def\citenamefont#1{#1}\fi
\expandafter\ifx\csname url\endcsname\relax
  \def\url#1{\texttt{#1}}\fi
\expandafter\ifx\csname urlprefix\endcsname\relax\def\urlprefix{URL }\fi
\providecommand{\bibinfo}[2]{#2}
\providecommand{\eprint}[2][]{\url{#2}}

\bibitem[{\citenamefont{Brown et~al.}(2004)\citenamefont{Brown, Grandchamp,
  Lee, and Rho}}]{Brown:2003wa}
\bibinfo{author}{\bibfnamefont{G.~E.} \bibnamefont{Brown}},
  \bibinfo{author}{\bibfnamefont{L.}~\bibnamefont{Grandchamp}},
  \bibinfo{author}{\bibfnamefont{C.-H.} \bibnamefont{Lee}}, \bibnamefont{and}
  \bibinfo{author}{\bibfnamefont{M.}~\bibnamefont{Rho}},
  \bibinfo{journal}{Phys. Rept.} \textbf{\bibinfo{volume}{391}},
  \bibinfo{pages}{353} (\bibinfo{year}{2004}), \eprint{hep-ph/0308147}.

\bibitem[{\citenamefont{Witten}(1984)}]{PhysRevD.30.272}
\bibinfo{author}{\bibfnamefont{E.}~\bibnamefont{Witten}},
  \bibinfo{journal}{Phys. Rev. D} \textbf{\bibinfo{volume}{30}},
  \bibinfo{pages}{272} (\bibinfo{year}{1984}).

\bibitem[{\citenamefont{Pisarski and Wilczek}(1984)}]{Pisarski:1983ms}
\bibinfo{author}{\bibfnamefont{R.~D.} \bibnamefont{Pisarski}} \bibnamefont{and}
  \bibinfo{author}{\bibfnamefont{F.}~\bibnamefont{Wilczek}},
  \bibinfo{journal}{Phys. Rev.} \textbf{\bibinfo{volume}{D29}},
  \bibinfo{pages}{338} (\bibinfo{year}{1984}).

\bibitem[{\citenamefont{D'Elia et~al.}(2006)\citenamefont{D'Elia, Di~Giacomo,
  and Pica}}]{D'Elia:2005sy}
\bibinfo{author}{\bibfnamefont{M.}~\bibnamefont{D'Elia}},
  \bibinfo{author}{\bibfnamefont{A.}~\bibnamefont{Di~Giacomo}},
  \bibnamefont{and} \bibinfo{author}{\bibfnamefont{C.}~\bibnamefont{Pica}},
  \bibinfo{journal}{PoS} \textbf{\bibinfo{volume}{LAT2005}},
  \bibinfo{pages}{158} (\bibinfo{year}{2006}), \eprint{hep-lat/0510012}.

\bibitem[{\citenamefont{Engels et~al.}(2006)\citenamefont{Engels, Holtmann, and
  Schulze}}]{Engels:2005rr}
\bibinfo{author}{\bibfnamefont{J.}~\bibnamefont{Engels}},
  \bibinfo{author}{\bibfnamefont{S.}~\bibnamefont{Holtmann}}, \bibnamefont{and}
  \bibinfo{author}{\bibfnamefont{T.}~\bibnamefont{Schulze}},
  \bibinfo{journal}{PoS} \textbf{\bibinfo{volume}{LAT2005}},
  \bibinfo{pages}{148} (\bibinfo{year}{2006}), \eprint{hep-lat/0509010}.

\bibitem[{\citenamefont{de~Forcrand and Philipsen}(2007)}]{deForcrand:2006pv}
\bibinfo{author}{\bibfnamefont{P.}~\bibnamefont{de~Forcrand}} \bibnamefont{and}
  \bibinfo{author}{\bibfnamefont{O.}~\bibnamefont{Philipsen}},
  \bibinfo{journal}{JHEP} \textbf{\bibinfo{volume}{01}}, \bibinfo{pages}{077}
  (\bibinfo{year}{2007}), \eprint{hep-lat/0607017}.

\bibitem[{\citenamefont{Laermann and Philipsen}(2003)}]{Laermann:2003cv}
\bibinfo{author}{\bibfnamefont{E.}~\bibnamefont{Laermann}} \bibnamefont{and}
  \bibinfo{author}{\bibfnamefont{O.}~\bibnamefont{Philipsen}},
  \bibinfo{journal}{Ann. Rev. Nucl. Part. Sci.} \textbf{\bibinfo{volume}{53}},
  \bibinfo{pages}{163} (\bibinfo{year}{2003}), \eprint{hep-ph/0303042}.

\bibitem[{\citenamefont{Heller}(2006)}]{Heller:2006ub}
\bibinfo{author}{\bibfnamefont{U.~M.} \bibnamefont{Heller}},
  \bibinfo{journal}{PoS} \textbf{\bibinfo{volume}{LAT2006}},
  \bibinfo{pages}{011} (\bibinfo{year}{2006}), \eprint{hep-lat/0610114}.

\bibitem[{\citenamefont{Parisen~Toldin
  et~al.}(2003)\citenamefont{Parisen~Toldin, Pelissetto, and
  Vicari}}]{ParisenToldin:2003hq}
\bibinfo{author}{\bibfnamefont{F.}~\bibnamefont{Parisen~Toldin}},
  \bibinfo{author}{\bibfnamefont{A.}~\bibnamefont{Pelissetto}},
  \bibnamefont{and} \bibinfo{author}{\bibfnamefont{E.}~\bibnamefont{Vicari}},
  \bibinfo{journal}{JHEP} \textbf{\bibinfo{volume}{07}}, \bibinfo{pages}{029}
  (\bibinfo{year}{2003}), \eprint{hep-ph/0305264}.

\bibitem[{\citenamefont{Lenaghan}(2001)}]{Lenaghan:2000kr}
\bibinfo{author}{\bibfnamefont{J.~T.} \bibnamefont{Lenaghan}},
  \bibinfo{journal}{Phys. Rev.} \textbf{\bibinfo{volume}{D63}},
  \bibinfo{pages}{037901} (\bibinfo{year}{2001}), \eprint{hep-ph/0005330}.

\bibitem[{\citenamefont{Marchi and Meggiolaro}(2003)}]{Marchi:2003wq}
\bibinfo{author}{\bibfnamefont{M.}~\bibnamefont{Marchi}} \bibnamefont{and}
  \bibinfo{author}{\bibfnamefont{E.}~\bibnamefont{Meggiolaro}},
  \bibinfo{journal}{Nucl. Phys.} \textbf{\bibinfo{volume}{B665}},
  \bibinfo{pages}{425} (\bibinfo{year}{2003}), \eprint{hep-ph/0301247}.

\bibitem[{\citenamefont{Basile et~al.}(2006)\citenamefont{Basile, Pelissetto,
  and Vicari}}]{Basile:2005hw}
\bibinfo{author}{\bibfnamefont{F.}~\bibnamefont{Basile}},
  \bibinfo{author}{\bibfnamefont{A.}~\bibnamefont{Pelissetto}},
  \bibnamefont{and} \bibinfo{author}{\bibfnamefont{E.}~\bibnamefont{Vicari}},
  \bibinfo{journal}{PoS} \textbf{\bibinfo{volume}{LAT2005}},
  \bibinfo{pages}{199} (\bibinfo{year}{2006}), \eprint{hep-lat/0509018}.

\bibitem[{\citenamefont{Adams and Chandrasekharan}(2003)}]{Adams:2003cc}
\bibinfo{author}{\bibfnamefont{D.~H.} \bibnamefont{Adams}} \bibnamefont{and}
  \bibinfo{author}{\bibfnamefont{S.}~\bibnamefont{Chandrasekharan}},
  \bibinfo{journal}{Nucl. Phys.} \textbf{\bibinfo{volume}{B662}},
  \bibinfo{pages}{220} (\bibinfo{year}{2003}), \eprint{hep-lat/0303003}.

\bibitem[{\citenamefont{Chandrasekharan and
  Jiang}(2006)}]{Chandrasekharan:2006tz}
\bibinfo{author}{\bibfnamefont{S.}~\bibnamefont{Chandrasekharan}}
  \bibnamefont{and} \bibinfo{author}{\bibfnamefont{F.-J.} \bibnamefont{Jiang}},
  \bibinfo{journal}{Phys. Rev.} \textbf{\bibinfo{volume}{D74}},
  \bibinfo{pages}{014506} (\bibinfo{year}{2006}), \eprint{hep-lat/0602031}.

\bibitem[{\citenamefont{Cecile and Chandrasekharan}(2006)}]{Cecile:2006}
\bibinfo{author}{\bibfnamefont{D.~J.} \bibnamefont{Cecile}} \bibnamefont{and}
  \bibinfo{author}{\bibfnamefont{S.}~\bibnamefont{Chandrasekharan}}
  (\bibinfo{year}{2006}), \bibinfo{note}{in preparation}.

\bibitem[{\citenamefont{Chandrasekharan and
  Mehta}(2006)}]{Chandrasekharan:2006zq}
\bibinfo{author}{\bibfnamefont{S.}~\bibnamefont{Chandrasekharan}}
  \bibnamefont{and} \bibinfo{author}{\bibfnamefont{A.~C.} \bibnamefont{Mehta}}
  (\bibinfo{year}{2006}), \eprint{hep-lat/0611025}.

\bibitem[{\citenamefont{Cecile}(2006)}]{Cecile:2006zr}
\bibinfo{author}{\bibfnamefont{D.~J.} \bibnamefont{Cecile}}
  (\bibinfo{year}{2006}), \eprint{hep-lat/0611026}.

\bibitem[{\citenamefont{Amit}(2005)}]{Amit}
\bibinfo{author}{\bibfnamefont{D.~J.} \bibnamefont{Amit}},
  \emph{\bibinfo{title}{Field Theory, the renormalization group and Critical
  Phenomena}} (\bibinfo{publisher}{World Scientific},
  \bibinfo{address}{Singapore}, \bibinfo{year}{2005}).

\bibitem[{\citenamefont{Chandrasekharan et~al.}(1999)}]{Chandrasekharan:1998yx}
\bibinfo{author}{\bibfnamefont{S.}~\bibnamefont{Chandrasekharan}}
  \bibnamefont{et~al.}, \bibinfo{journal}{Phys. Rev. Lett.}
  \textbf{\bibinfo{volume}{82}}, \bibinfo{pages}{2463} (\bibinfo{year}{1999}),
  \eprint{hep-lat/9807018}.

\end{thebibliography}

\end{document}